# Laser Spectroscopic Measurement of Helium Isotope Ratios

L.-B. Wang

*Physics Division, Argonne National Laboratory, Argonne, USA; Physics Department, University of Illinois at Urbana-Champaign, Urbana, USA*

P. Mueller, R. J. Holt, Z.-T. Lu,* T. P. O'Connor

*Physics Division, Argonne National Laboratory, Argonne, USA*

Y. Sano

*Ocean Research Institute, The University of Tokyo, Nakanoku, Japan*

N. C. Sturchio

*Department of Earth and Environmental Sciences, University of Illinois at Chicago, Chicago, USA*

**Abstract.** A sensitive laser spectroscopic method has been applied to the quantitative determination of the isotope ratio of helium at the level of $^3$He/$^4$He = $10^{-7}$ - $10^{-5}$. The resonant absorption of 1083 nm laser light by the metastable $^3$He atoms in a discharge cell was measured with the frequency modulation saturation spectroscopy technique while the abundance of $^4$He was measured by a direct absorption technique. The results on three different samples extracted from the atmosphere and commercial helium gas were in good agreement with values obtained with mass spectrometry. The achieved 3σ detection limit of $^3$He in helium is $4 \times 10^{-9}$. This demonstration required a 200 μL STP sample of He. The sensitivity can be further improved, and the required sample size reduced, by several orders of magnitude with the addition of cavity enhanced spectroscopy.

* Corresponding author. Email: lu@anl.gov; URL: www-mep.phy.anl.gov/atta/.

The ratio of the stable He isotopes in the Earth's atmosphere ($R_{atm}$) is $^3$He/$^4$He ~ $1.4 \times 10^{-6}$ (Mamyrin et al., 1970). This value represents a steady-state mixture of He released from the Earth into the atmosphere where it has a residence time of only about 1 million years before escaping into space (Torgerson, 1989). The He isotope ratio in primitive solar system material and in the lunar soil is relatively high at 120-330 $R_{atm}$ (Pepin and Porcelli, 2002). Helium in mantle-derived volcanic rocks and emitted from volcanic fumaroles has a He isotope ratio about 8 $R_{atm}$ (Craig et al., 1978), whereas radiogenic He in continental rocks has a He isotope ratio about 0.01 $R_{atm}$ (Morrison and Pine, 1955; Mamyrin and Tolstikhin, 1984). Thus, the range in He isotope ratios of natural materials is about 0.01 to 330 $R_{atm}$, or $^3$He/$^4$He = $10^{-8}$ to $5 \times 10^{-4}$. Measurements of He isotope ratios have a number of important geoscientific applications. For example, they are used to trace volatile emanations from the Earth's mantle, such as those occurring in volcanic areas (Craig et al., 1978); to trace the input of micrometeoritic and cometary material into ocean sediments through time (Marcantonio et al., 1996); to determine the residence time of young groundwaters via the $^3$H-$^3$He method (Schlosser et al., 1998); to determine the cosmic ray exposure ages of surface rocks (Kurz, 1986); and to prospect the lunar surface for $^3$He "ore" to fuel fusion energy reactors (Wittenberg et al., 1968).

A typical measurement of the He isotope ratio in natural materials requires extraction and purification of the He, followed by introduction of about 5 nL of He into a specially designed magnetic sector mass spectrometer. Approximately $1 \times 10^7$ $^3$He atoms are required to obtain a precise measurement of the $^3$He/$^4$He ratio (Clarke et al., 1976). Alternatively, Kudryavtsev et al. (1989) explored a collinear resonance ionization method and performed measurements of $^3$He/$^4$He ratios down to the $10^{-8}$ level. However, their setup was rather complex. It consisted of multiple lasers at different wavelengths and a large atomic beam system. The purpose of this communication is to demonstrate a new spectroscopic method for the measurement of He isotope ratios. Since the spectroscopic instrument is significantly smaller (more compact), lighter, and less expensive to purchase and operate than a conventional noble gas mass spectrometer, it would be ideal to use in field studies related to earthquake prediction (Sano et al., 1998), volcanic hydrothermal systems (Sturchio et al., 1993) and chemical oceanography (Lupton and Craig, 1981).

In this work we use a laser spectroscopic technique and take advantage of the isotope shift between $^3$He and $^4$He to separately probe the two isotopes. Unfortunately, laser excitation of ground-level helium atoms is at present only realized with pulsed and sophisticated vacuum-ultraviolet laser systems. Instead, we choose to excite the metastable helium atoms via the $2^3S \rightarrow 2^3P$ transition for the resonant absorption measurements because it has a reasonable transition strength and is accessible with commercially available lasers at 1083 nm (Zhao et al., 1991). The resonant absorption peaks of the two isotopes are 65 GHz apart, well separated compared to the Doppler width of 2 GHz and the natural linewidth of 1.6 MHz (Fig. 1). The resonant absorption signal of the rare isotope $^3$He, $S_3$, is measured using frequency-modulation (FM) saturation spectroscopy. This method was first demonstrated by Bjoklund (1980) and Hall et al. (1981). The variations of this method have previously been used to detect molecular species at the parts-per-million level (e.g. Wang et al., 1995, Petrov et al., 1997, and Modungno et al., 1998), but to our knowledge it has not been used to detect rare isotopes at the comparable or lower level. The resonant absorption signal of the abundant $^4$He, $S_4$, is measured simply with a single laser beam. In this work we show that the signal ratio of these two measurements, $S_3/S_4$, is indeed proportional to the isotope ratio, $^3$He/$^4$He. Once the signal ratio is calibrated with a standard sample of known isotope ratio, measurements on unknown samples can then be used to determine the unknown isotope ratios.

The laser spectroscopic measurements are performed at Argonne National Laboratory on three different samples in which the isotopic abundance of $^3$He is also measured at the University of Tokyo with the well-established mass spectrometry method. Sample #1 is extracted from the surrounding air in the laboratory with a sorption pump cooled down to 80 K by liquid nitrogen, which effectively absorbs all major gases in air except neon, helium and hydrogen, whose partial pressures in air are 13.8 mTorr, 4.0 mTorr and 0.4 mTorr, respectively. After the sorption pump has reached equilibrium the remaining gas is compressed with a turbopump into an evacuated cell for laser spectroscopy work. In the cell, the chemically active gases such as hydrogen and water are absorbed by a getter pump, and their partial pressures are reduced to less than 1% of the total pressure, leaving only neon and helium with the pressure ratio of approximately 3.5:1 as measured with a residual gas analyzer. The gas purity of the sample proves to be critical as any impurities except neon can quench the metastable state of helium upon collisions. Sample #2 is a commercial helium gas (99.995% pure, AGA UN1046) extracted from natural gas. Sample #3 is a relatively $^3$He-enriched helium produced earlier by a collaboration of four noble gas laboratories in Japan (Matsuda et al., 2002).

The gas sample is contained in a 1 m long and 2.5 cm diameter glass cell, around which a RF-driven discharge is used to populate the metastable 1s2s $^3S_1$ level via electron-impact collisions. Tests show that the amplitude of the $^3$He signal varies by a factor of two in the pressure range of 50-400 mTorr. For an optimum $^3$He signal the pressure should be approximately 200 mTorr, at which an estimated metastable population of approximately $1 \times 10^{-4}$ relative to groundlevel helium atoms is maintained. The minimum sample size needed to fill this cell to the operation pressure is 200 $\mu$L STP. Although the sample size can be decreased by one order of magnitude with a better design of the vacuum system, it is still much larger than that required by analyses with mass spectrometry. A laser system consisting of two grating stabilized diode-lasers and a fiber amplifier provides the required laser power of 500 mW and the single-mode laser frequency with long-term stability and scan control of better than 1 MHz.

The setup of FM saturation spectroscopy of $^3$He is shown in Figure 2. This approach is described in detail in Hall et al. (1981). The probe laser beam is phase-modulated at 36 MHz with an electro-optical modulator, then sent through the glass cell that contains the gas sample with a power of 8 mW and a diameter of 1 cm, and finally focused onto a fast InGaAs-photodiode detector. The pump beam is frequency shifted by 2 MHz and chopped at 45 kHz with two acousto-optical modulators, then sent through the long cell in the opposite direction with a power of 50 mW and a diameter of 1 cm. The signal from the photodiode is first demodulated at 36 MHz with a RF frequency mixer and then at 45 kHz with a lock-in amplifier, whose output is recorded as data. When the laser frequency is scanned over the resonance, a signal with the shape of the derivative of a Lorentzian is observed (Fig. 3). Details of the line shape calculation can be found in Supplee et al. (1994). For natural samples with the $^3$He/$^4$He ratio ranging from 0.01 $R_{atm}$ to 330 $R_{atm}$, the corresponding resonant absorption in this setup can be set to a range from $1\times10^{-6}$ to $3\times10^{-2}$, over which the signal size is expected to be linearly proportional to the number of metastable $^3$He atoms resonantly interacting with the laser beams.

The absorption signal, however, is affected by the metastable fraction which depends on discharge intensity, gas pressure, and the chemical composition of the sample. In order to reduce the potential systematic errors due to the changes in the metastable fraction, the abundance of the metastable $^4$He is measured for normalization. The abundance of metastable $^4$He is determined in a single laser beam absorption measurement. Using the same setup, the level of absorption of the probe beam is measured directly at the DC-output of the photodiode without any modulation techniques. If the laser frequency is tuned to the resonance of the $2^3S_1 \rightarrow 2^3P_0$ transition of $^4$He, the probe beam is totally absorbed within the Doppler width of 2 GHz. Instead, the laser frequency is tuned to 1.5 GHz away from resonance and the discharge intensity is adjusted by controlling the RF power in order to achieve a condition where the fractional absorption is 50%. The discharge intensity is preserved between the $^4$He and $^3$He measurements so that the $^3$He measurement is effectively normalized by the $^4$He measurement.

We find the following techniques effective in reducing detection noise: the probe beam is focused and passed through a 50 μm pinhole to eliminate the scattering light from the pump beam; a 60 dB optical isolator is installed in front of the fiber amplifier output to reduce optical feedback caused by the reflection from the front facet of the photodetector; a slowly drifting interference effect between the pump and probe beam is minimized by shifting the pump beam frequency by 2 MHz and by slightly misaligning the two beam paths. The achieved detection noise is approximately 5 times the shot noise limit. At present, the dominant noise is due to the fluctuation in the power of both the diode laser and the fiber amplifier caused by the pickup of the intense RF power needed to drive the discharge. More thorough and effective RF shielding should help mitigate this problem.

Figure 3 shows a typical scan over the $^3$He resonance. The total scan range is 320 MHz and the scan is accumulated over 28 minutes. The stronger peak is due to the $2^3S_{1, F=3/2} \rightarrow 2^3P_{2, F=5/2}$ transition and the weaker peak, 221 MHz away, is due to the $2^3S_{1, F=1/2} \rightarrow 2^3P_{2, F=3/2}$ transition. The signal-to-noise ratio is 200 with sample #2 whose isotope ratio ($^3$He/$^4$He) is $2.8 \times 10^{-7}$. The corresponding $3\sigma$ detection limit is then $4 \times 10^{-9}$. Our measurements are in good agreement with the mass spectrometry values as shown in Table 1 and Figure 4. The uncertainties of our measurements are dominated by system instability caused by the change of discharge condition and the laser power drift between the measurements on $^3$He and $^4$He. These effects can be reduced by a better controlled and more stable system and by implementing a more frequent switch between the measurements on $^3$He and $^4$He.

In conclusion, we have demonstrated that the isotopic abundance of $^3$He in natural He samples can be determined quantitatively with a laser absorption technique. The achieved $3\sigma$ detection limit on the $^3$He/$^4$He isotope ratio is $4\times10^{-9}$ with the present system containing a 200 $\mu$L STP helium sample. We believe that the sensitivity of this method can be improved and the sample size decreased by several orders of magnitude. Near-term improvements such as a more thorough RF shielding can reduce the detection noise by a factor of 5, down to the shot noise limit. Furthermore, the cell can be installed in a multi-pass Fabry-Perot cavity to increase the effective absorption length by as much as a factor of $1 \times 10^5$ (Ma et al., 1999). In the distant future, when continuous-wave and narrow bandwidth lasers at 58 nm may become available, laser spectroscopy would then be performed on groundlevel He atoms rather than metastable He atoms, and the detection sensitivity would be further improved.

**Acknowledgments.** We thank J.P. Schiffer for stimulating discussions, K. Bailey for general technical support, and J. Gregar for his artful glass work. This work is supported by the U.S. Department of Energy, Nuclear Physics Division, under contract W-31-109-ENG-38.

Mailing address: L.-B. Wang, P. Mueller, R. J. Holt, Z.-T. Lu, T. P. O'Connor, *Physics Division, Argonne National Laboratory, Argonne, IL 60439, USA.* Y. Sano, *Ocean Research Institute, The University of Tokyo, Nakanoku, Tokyo 164-8639, Japan.* N. Sturchio, *Department of Earth and Environmental Sciences, University of Illinois at Chicago, Chicago, Illinois 60637, USA*

**Fig. 1.** Energy level diagram (not to scale) for the $2^3S$ and $2^3P$ states of $^3He$ and $^4He$ (Zhao et al., 1991). The frequency differences between energy levels are indicated in the unit of MHz. The dotted lines show the transitions used for the measurements. $f_4 = f_3 + 65$ GHz.

**Fig. 2.** Experimental set-up used for frequency-modulation saturation spectroscopy. The glass cell is custom-made; all the other components are commercially available. AOM, acousto-optical modulator; DL, diode laser; EOM, electro-optical modulator; FA, fiber amplifier; FG, function generator; PBS, polarization beam splitter; PD, photodiode detector; RF, RF signal generator.

**Fig. 3.** FM spectroscopy signal of $^3He$. The stronger peak is due to the $2^3S_{1, F=3/2} \to 2^3P_{2, F=5/2}$ transition and the weaker peak, 221 MHz away, is due to the $2^3S_{1, F=1/2} \to 2^3P_{2, F=3/2}$ transition.

**Fig. 4.** $^3He/^4He$ isotope ratios, laser spectroscopy vs. mass spectrometry.

**Table 1.** Laser spectroscopy vs. mass spectrometry

| Heading 1 | $^3He$ FM signal | $^3He / ^4He$ laser spectroscopy[*] | $^3He / ^4He$ mass spectrometry[§] |
|---|---|---|---|
| 1. helium from air | 3.61 ± 0.31 | (1.4 ± 0.1)×10$^{-6}$ | (1.39 ± 0.01)×10$^{-6}$ |
| 2. helium gas bottle | 0.72 ± 0.12 | (2.8 ± 0.5)×10$^{-7}$ | (2.81 ± 0.28)×10$^{-7}$ |
| 3. standard sample | 75.4 ± 9.9 | (2.9 ± 0.4)×10$^{-5}$ | (2.87 ± 0.14)×10$^{-5}$ |

[*] The laser method is calibrated using the atmospheric isotope ratio $^3He/^4He = 1.39 \times 10^{-6}$.
[§] The mass spectrometry value for sample #1 is derived from Mamyrin et al. (1970) and Clarke et al. (1976). The mass spectrometry value for sample #3 is derived from measurements by four noble gas laboratories in Japan (Matsuda et al., 2002).

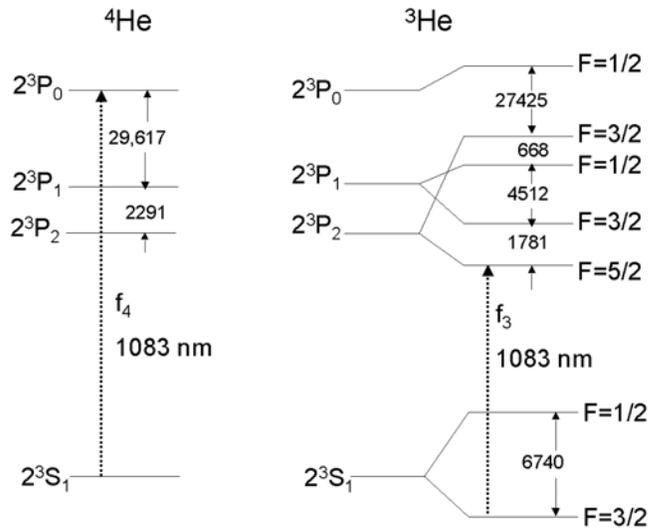

**Fig. 1.** Energy level diagram (not to scale) for the $2^3S$ and $2^3P$ states of $^3He$ and $^4He$ (Zhao et al., 1991). The frequency differences between energy levels are indicated in the unit of MHz. The dotted lines show the transitions used for the measurements. $f_4 = f_3 + 65$ GHz.

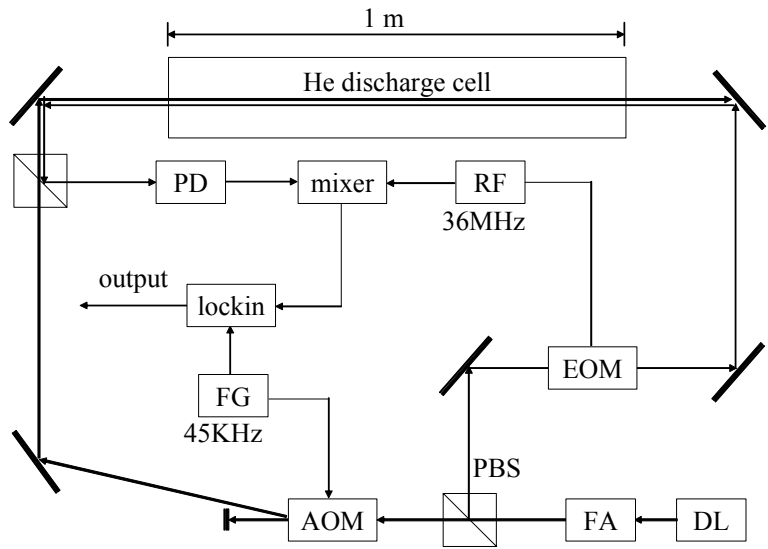

**Fig. 2.** Experimental set-up used for frequency-modulation saturation spectroscopy. The glass cell is custom-made; all the other components are commercially available. AOM, acousto-optical modulator; DL, diode laser; EOM, electro-optical modulator; FA, fiber amplifier; FG, function generator; PBS, polarization beam splitter; PD, photodiode detector; RF, RF signal generator.

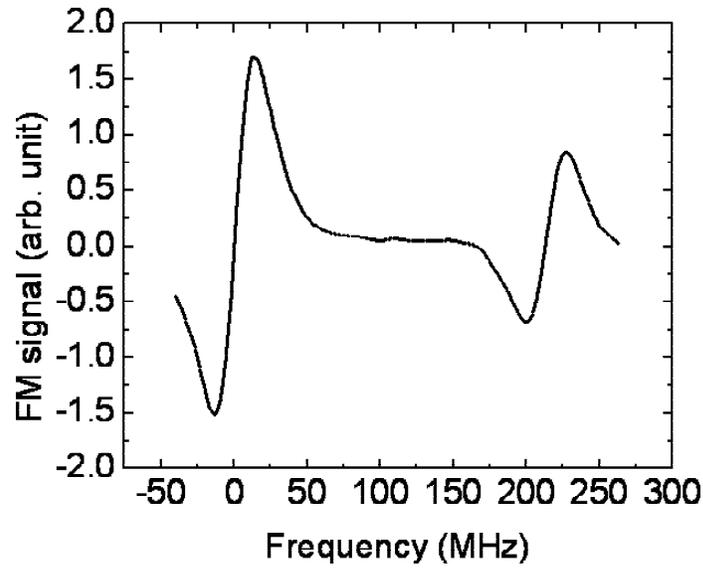

**Fig. 3.** FM spectroscopy signal of $^3$He. The stronger peak is due to the $2^3S_{1, F=3/2} \rightarrow 2^3P_{2, F=5/2}$ transition and the weaker peak, 221 MHz away, is due to the $2^3S_{1, F=1/2} \rightarrow 2^3P_{2, F=3/2}$ transition.

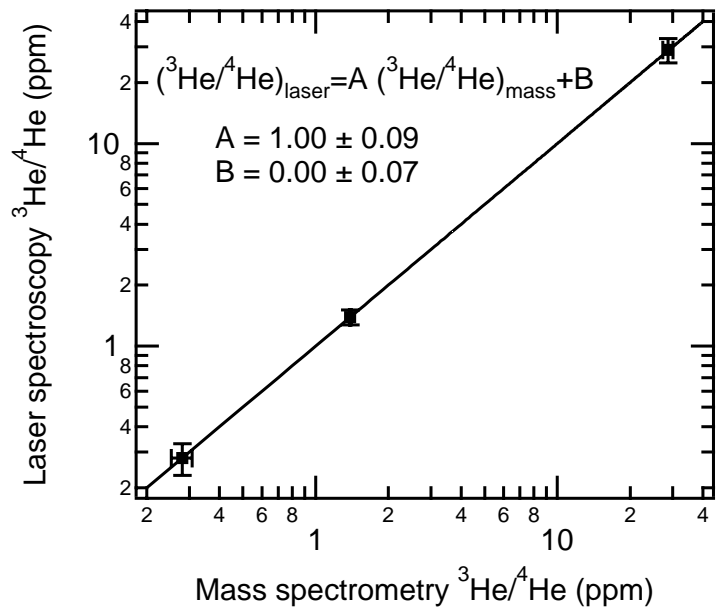

**Fig. 4.** $^3$He/$^4$He isotope ratios, laser spectroscopy vs. mass spectrometry.